\documentclass{article}

\usepackage[numbers, compress]{natbib}

 \usepackage[final]{AdvML_Frontiers_2024}

\usepackage[utf8]{inputenc} 
\usepackage[T1]{fontenc}    
\usepackage{hyperref}       
\usepackage{url}            
\usepackage{booktabs}       
\usepackage{amsfonts}       
\usepackage{nicefrac}       
\usepackage{microtype}      
\usepackage{xcolor}         
\usepackage{algpseudocode}
\usepackage{algorithm}
\usepackage{amsmath,amsfonts, amssymb}
\usepackage{graphicx}

\title{Robustness of Practical Perceptual Hashing Algorithms to Hash-Evasion and Hash-Inversion Attacks}

\author{
Jordan Madden, Moxanki Bhavsar, Lhamo Dorje, and Xiaohua Li
\\
  Dept. of ECE, 
  Binghamton University\\
  Binghamton, NY 13902 \\
  \texttt{\{jmadden2, mbhavsa2, ldorje1, xli\}@binghamton.edu} \\
}

\begin{document}

\maketitle

\begin{abstract}
Perceptual hashing algorithms (PHAs) are widely used for identifying illegal online content and are thus integral to various sensitive applications. However, due to their hasty deployment in real-world scenarios, their adversarial security has not been thoroughly evaluated. This paper assesses the security of three widely utilized PHAs—PhotoDNA, PDQ, and NeuralHash—against hash-evasion and hash-inversion attacks. Contrary to existing literature, our findings indicate that these PHAs demonstrate significant robustness against such attacks. We provide an explanation for these differing results, highlighting that the inherent robustness is partially due to the random hash variations characteristic of PHAs. Additionally, we propose a defense method that enhances security by intentionally introducing perturbations into the hashes.


\end{abstract}


\section{Introduction} 
\label{sec1}

Perceptual hashing, also known as robust hashing, is a content fingerprinting technique that generates similar binary hashes for perceptually similar contents and different hashes for different contents \cite{barni2023information}. Perceptual hashing is useful in various applications such as content moderation, authentication, and detection.
Numerous perceptual hash algorithms (PHAs) have been developed or deployed for practical applications \cite{farid2021overview}. Early examples include the algorithm used on YouTube to identify copyrighted material \cite{sivic2003video} and algorithms used in online image search engines of Google, Microsoft Bing, etc \cite{hao2021s}.

Recently, PHAs have gained more critical importance, particularly in responding to the requiremnt of reporting illicit content. For instance, in cases involving Child Sexual Abuse Media (CSAM), US media service providers are legally obliged to examine suspected illegal content and report it to a ``Cyber Tip Line'' operated by the National Center for Missing and Exploited Children (NCMEC). Microsoft's PhotoDNA and Meta's PDQ were immediately deployed for this purpose 
\cite{logan2022citizen}\cite{ith2015microsoft}\cite{davis2019open}. 

With a growing emphasis on user privacy, technologies like end-to-end encryption have become prevalent. This shift has led to the deployment of PHAs on the client side before data encryption \cite{kulshrestha2021identifying}. For example, Apple has proposed to deploy NeuralHash on all its iOS devices \cite{cobbe2021data}. 
However, this client-side deployment has raised significant concerns among the client users about their data privacy, which inadvertently arouses a widespread criticism regarding the security and privacy-preserving capabilities of PHAs \cite{abelson2024bugs}. Criticism has been further fueled by recent studies highlighting the vulnerability of PHAs to adversarial attacks and the potential for information leakage through hash bits \cite{hao2021s}\cite{mckeown2023hamming}\cite{Athalye2021inverting}.

Given the critical role PHAs play in sensitive applications, it is essential to ensure their security. Unfortunately, PHAs are often deployed hastily without sufficient security evaluation. 
This paper systematically assesses the adversarial robustness of three widely used PHAs—PhotoDNA, PDQ, and NeuralHash—against two types of attacks: hash-evasion attacks, aimed at enabling illicit content to evade detection, and hash-inversion attacks, aimed at reconstructing the original secret content from hashes. To support this evaluation, we introduce two new attack algorithms: a query-efficient black-box adversarial attack algorithm and a data-efficient hash inversion algorithm.

The major contributions of this paper are as follows.
\begin{itemize}

\item This study demonstrates that PHAs exhibit notable robustness against blackbox adversarial evasion attacks when realistic constraints on image distortion, query budget, and hash difference are applied. This finding challenges previous studies that overlooked these constraints, leading to unjustified criticisms of PHAs.

\item This paper confirms the sufficient privacy-preserving capabilities of hash bits, countering claims from some existing studies that suggested hash bits disclose significant information about the content of the original images. The key point is how diverse the images are.

\item This research reveals that the security of PHAs is partially due to the unique property of random hash variations. Based on this observation, a defense method is proposed to enhance security by intentionally randomizing hash bits. 

\end{itemize}

The paper is organized as follows: Section \ref{sec2} provides a literature review, Section \ref{sec3} formulates the attack and defense algorithms, Section \ref{sec4} presents the experiment results, and Section \ref{sec5} concludes the paper. More detailed experiment data can be found in the Appendix. The source code of the experiments, including ways to access binary executables of the PHAs,  can be found at \emph{ \url{ https://github.com/neddamj/nnhash}}.

\section{Literature Review}
\label{sec2}

Among the numerous PHA's \cite{farid2021overview}, this paper focuses on the following three major ones due to their widespread deployment in practice:

\begin{itemize}
\item {\bf PhotoDNA}, released by Microsoft in 2009, is used by over 70 companies including Cloudflare and Dropbox \cite{ith2015microsoft}\cite{burgett2014photodna}\cite{arthur2013twitter}. It generates an 1152-bit hash for each input image;

\item {\bf PDQ}, launched by Meta in 2019 \cite{davis2019open} and heavily inspired by pHash \cite{farid2021overview}\cite{klinger2013phash}, produces a 256-bit hash for each input image.

\item {\bf NeuralHash}, released by Apple in 2021 with the goal of identifying CSAM \cite{cobbe2021data}, creates a more concise 96-bit hash for each input image.

\end{itemize}

The adversarial robustness of PhotoDNA, PDQ, and NeuralHash to either hash-evasion or hash-inversion attacks has not been thoroughly investigated, primarily due to their relative novelty and the absence of publicly available source code. For hash-evasion, \cite{struppek2022learning} was the first to explore the robustness of NeuralHash, revealing its vulnerability to whitebox attacks, assuming the adversary has full access to the PHA. Additionally, \cite{jain2022adversarial} examined PDQ and demonstrated how adversarial images could be generated in the whitebox context. Furthermore, \cite{bhatia2022exploiting} utilized linear interpolation between two images to create adversarial examples.

Regarding blackbox hash-evasion attacks, where the adversary only has access to the input and output of the algorithm, \cite{jain2022adversarial} examined PDQ. It concluded that it was not robust against untargeted blackbox attacks. Similarly, \cite{prokos2023squint} evaluated both PhotoDNA and PDQ, finding that they lacked robustness against both targeted and untargeted attacks. While these studies are closely related to the current paper, they overlooked the importance of adversaries' constraints on image distortion, query budget, as well as hash difference levels. By all means, unconstrained blackbox attacks may always be successful, particularly in the untargeted setting.

Regarding hash-inversion attacks or privacy-preserving capabilities of PHAs, Microsoft initially claimed that the hashes generated by PhotoDNA were irreversible and could not be used to reconstruct the original images \cite{grimm2024putting}. However, a blog post \cite{Athalye2021inverting} challenged this assertion, suggesting that PhotoDNA hashes might contain enough information to synthesize images from the hashes. To the best of our knowledge, \cite{Athalye2021inverting} is the only work addressing the reconstruction of original images from hashes, and it specifically focuses on PhotoDNA, which uses long hashes.
Another method for investigating information leakage involves training classifiers to identify image content from hashes. Studies such as \cite{struppek2022learning} and \cite{bhatia2022exploiting} demonstrated that NeuralHash hashes could indeed reveal information about their source images, such as identifying the object's class. These findings contrast with those of \cite{nadeem2019privacy}, which examined PhotoDNA hashes and concluded that they were resistant to such classification attacks.

\section{Attack Algorithms and Defense Method}
\label{sec3}

In a perceptual hash system, a service provider scans a client’s unencrypted content, extracts hashes for each image using a perceptual hashing algorithm (PHA) and compares these hashes with a database of known illicit hashes. If a match is found, an alarm is raised or the unencrypted content is forwarded for further analysis \cite{kulshrestha2021identifying}. In this context, we consider two types of adversaries: the {\it client adversary}, who aims to evade hash matching, and the {\it service provider (or third-party) adversary}, who seeks to reconstruct the client’s original images from collected hashes.
This paper evaluates the security of three widely used PHAs—PhotoDNA, PDQ, and NeuralHash—against two specific types of attacks: blackbox hash-evasion attacks, which are employed by the client adversary, and hash-inversion attacks, which are utilized by the service provider adversary. Additionally, we propose a defense mechanism that can be integrated into the perceptual hashing system.


\subsection{A Query-Efficient Adversarial Blackbox Attack Algorithm for Hash Evasion}

This study focuses exclusively on blackbox adversarial attacks, rather than whitebox attacks, to reflect the practical scenario where PHA implementations are proprietary. In many cases, the details of PHA architectures and model weights are not publicly available due to their commercial nature. For example, Microsoft enforces non-disclosure agreements for PhotoDNA users, providing only an emulation of the algorithm. Similarly, although \cite{ygvar2021apple} released weights for the first-generation NeuralHash, Apple has since updated the algorithm, and the details of the latest version remain confidential.


Consider the original image ${\bf I}_0$ with the hash value ${\bf h}_0={\cal H}({\bf I}_0)$. The objective of the client adversary is to generate a very similar image ${\bf I}_{\rm adv}$ whose new hash ${\bf h}_{\rm adv} = {\cal H}({\bf I}_{\rm adv})$ is sufficiently different from ${\bf h}_0$ in untargeted attacks or sufficiently similar to some target hash ${\bf h}_{\rm tar}$ in targeted attacks. We use normalized Hamming distance $D_{h}({\bf h}_0, {\bf h}_{\rm adv})$ to evaluate the similarity of two hashes, 
and use per-pixel normalized $L_2$ norm $L_2({\bf I}_0, {\bf I}_{\rm adv})$ to evaluate the similarity of two images. 

Because the output of PHAs is hash bits, not logits or class decisions, we propose a joint soft-label hard-label attack (JSHA) algorithm that can drastically improve blackbox attack query efficiency. JSHA consists of two steps. The first step is to apply a soft-label blackbox attack to generate an initial adversarial image ${\bf I}_{\rm init}$ so that its hash ${\bf h}_{\rm init} = {\cal H}({\bf I}_{\rm init})$ is far away from ${\bf h}_0$ for untargeted attacks or close to ${\bf h}_{\rm tar}$ for targeted attacks. This step is formulated as the optimization
\begin{eqnarray}
   \max_{\delta} \;\;  D_h\left({\bf h}_0, {\cal H}({\bf I}_0 + \delta) \right) \; \text{(untargeted)}, \;\;\; \text{or,} \;\;\;            
    \min_{\delta} \;\;  D_h\left({\bf h}_{\rm tar}, {\cal H}({\bf I}_0 + \delta)   \right) \;  \text{(targeted)} 
    \label{eq30}
\end{eqnarray}
The optimization is conducted iteratively. In each iteration, the adversary queries the PHAs with its adversarial input ${\bf I}_0 + \delta$ to get the hash, and uses the hash to optimize the loss. After reaching the query number threshold or the hash distance threshold $D_0$,
this first step stops with output ${\bf I}_{\rm init}={\bf I}_0 + \delta$.

The second step applies a hard-label blackbox attack to reduce the image distortion while maintaining the hash distance obtained in the first step. Specifically, this step solves the following optimization
\begin{eqnarray}
    \min\limits_{{\bf I}_{\rm adv}} \;\; L_2({\bf I}_0, {\bf I}_{\rm adv}),
  s.t. \; D_h({\bf h}_0, {\cal H}({\bf I}_{\rm adv})) \geq D_0  \text{   (untargeted)},  D_h({\bf h}_0, {\cal H}({\bf I}_{\rm adv})) \leq D_0 \text{   (targeted)}   \label{eq50}
\end{eqnarray}
Starting from ${\bf I}_{\rm adv}={\bf I}_{\rm init}$, the adversary queries the PHAs to get the hash and uses the queried hash to optimize (\ref{eq50}) until the query number threshold or the threshold of $L_2({\bf I}_0, {\bf I}_{\rm adv})$ is reached. 

One of major advantages of this algorithm is high query efficiency. The first step is an unconstrained optimization and a large learning rate can be applied to get an initial adversarial sample quickly. In the second step, since we start from ${\bf I}_{\rm init}$ which is only a noisy version of ${\bf I}_0$, the convergence is fast. 






\begin{figure}[t]
\centerline{
\includegraphics[width=0.25\textwidth]{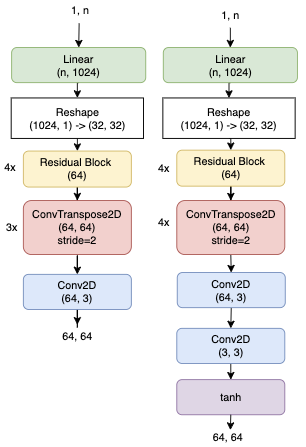}
\hspace{2cm}
\includegraphics[width=0.18\textwidth]{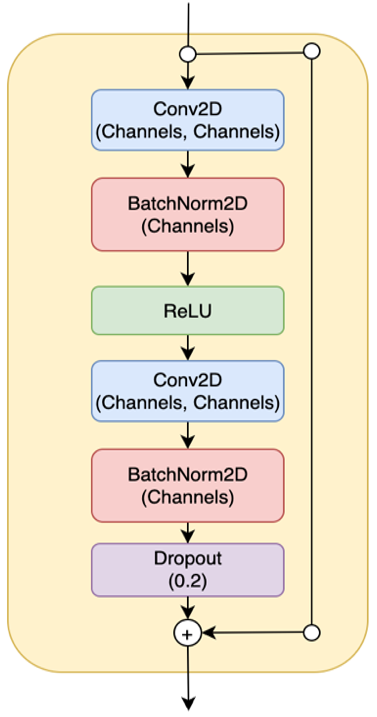}
}
\centerline{(a) \hspace{3.5cm} (b)}
\caption{(a) Hash inversion attack model (left: for Celeb Dataset. Right: for STL-10 dataset). (b) Architecture of residual block.} 
\label{fig:inversion_model}
\end{figure}

\subsection{A Data-Efficient Hash-Inversion Attack Algorithm}

GAN (generative adversarial network) was trained in \cite{Athalye2021inverting} to invert CelebA images. However, the adversary may need more data-efficient hash-inversion attack algorithms because the training data may not be abundant. For this, we has developed new and more data-efficient generative models to reconstruct the original input images from the hashes. We designed a decoder-style deep model, similar to the Fast Style Transfer \cite{johnson2016perceptual}, to convert the $N$-bit hashes to $(64, 64)$ images. This model is mainly a combination of residual blocks depicted in Fig. \ref{fig:inversion_model}(b), with fractionally-strided convolutions/deconvolutions for learned upsampling. The entire architecture is described in Fig. \ref{fig:inversion_model}(a), where the left network is for the MNIST and CelebA dataset while the right network is used for the STL-10 dataset. 

For a given training dataset, we first compute the hashes of each image offline and store the image-hash pairs. We then train the model outlined above to reconstruct the images as closely as possible by minimizing the mean-square-error (MSE) loss between the model output and the image from which the hash was generated. 



 \subsection{A Defense Method based on Random Hash Perturbation} 
 
The adversaries rely on gradient estimation to solve the optimization problems. It is well known that hash bits of PHAs change slightly and randomly with respect to slight changes in input images. Based on existing work on random perturbation defense \cite{qin2021random}\cite{aithal2022mitigating}, we have the following proposition:

{\it Proposition 1}. Gradient estimation based on queried hash is not reliable due to the random variation of hash bits.

An outline of the proof is in the Appendix A.1. The random variation of hash bits makes the adversaries' optimization struggle to converge, which will be demonstrated in the next section. Furthermore, considering this observation, we propose a defense method where the PHAs  randomly invert $q$-portion of the hash bits to guarantee sufficient hash randomness, where $0 \le q \le 1$.

\section{Experiment}
\label{sec4}

\subsection{Performance Metrics}

We use the attack success rate for assessing the robustness against blackbox adversarial hash-evasion attacks. Given a set of $M$ original images ${\bf I}_0^m$ with their corresponding hashes ${\bf h}_0^m$, the adversary generates adversarial images ${\bf I}_{\rm adv}^m$ with hashes ${\bf h}_{\rm adv}^m$. The attack success rate is defined as
\begin{equation}
  {\rm ASR}(p) = 
   \frac{1}{M}\sum_{m=1}^M \mathbb{I}\left(D_h({\bf h}_0^m, {\bf h}_{\rm adv}^m) \gtrless p)\right) \mathbb{I}\left(L_2({\bf I}_0^m, {\bf I}_{\rm adv}^m) < \theta \right)     \label{eq80}
\end{equation}
where $\mathbb{I}(\cdot)$ is the indicator function, $p\in [0, 1]$ is the pre-set hash distance threshold and $\theta$ is the pre-set image distortion threshold. For the operator ``$\gtrless$'', $>$ is used for untargeted attacks, $<$ is for targeted attacks. Small or zero ASR($p$) means a PHA is robust to an attack.

For the thresholds, we set $\theta=0.1$. On the other hand,
we should apply large enough $p$ to evaluate untargeted attacks because ${\bf h}_{\rm adv}$ should be different from ${\bf h}_0$. 
It was found in \cite{mckeown2023hamming} that most pairs of distinct images had $D_h({\bf h}_0, {\bf h}_1)>0.3$. Therefore, we primarily use $p=0.3$ in (\ref{eq80}) to determine the success of untargeted attacks. In contrast, we should use small enough $p$ for targeted attacks to make sure ${\bf h}_{\rm adv}$ is sufficiently close to ${\bf h}_{\rm tar}$. Prior studies \cite{mckeown2023hamming} indicate that many normal image edits would change an image's hashes to the level of $D_h({\bf h}_0, {\bf h}_1) < 0.1$. Therefore, we mainly use $p=0.1$ in (\ref{eq80}) for assessing the success of targeted attacks. This important parameter $p$ was unfortunately neglected by most existing adversarial robustness studies.

For hash-inversion attacks, we use the $L_2$ distortion, the SSIM (Structural Similarity Index Measure) \cite{hore2010image}, and the LPIPS (Learned Perceptual Image Patch Similarity) \cite{zhang2018unreasonable} between the reconstructed image and the true image to evaluate the performance. No ASR or query budget is involved.






\subsection{Experiment on Hash-Evasion Attack and Defense}

In implementing our proposed JSHA algorithm, we modified three different soft-label attack algorithms: SimBA \cite{guo2019simple}, NES \cite{ilyas2018black}, and ZOsignSGD \cite{liu2018signsgd}, to use in Step One, and modified the hard-label attack algorithm HSJA \cite{chen2020hopskipjumpattack} to use in Step Two. For instance, ``SimBA+HSJA'' means our JSHA with a modified SimBA as Step One and a modified HSJA as Step Two. Modifications included removing/changing constraints and adjusting hyper-parameters, etc. 
We compared the proposed JSHA with the direct application of the original SimBA, NES, and ZOsignSGD algorithms. We also compared it with existing works \cite{jain2022adversarial} and \cite{prokos2023squint}. 

The experiments were conducted on 100 randomly selected images from the ImageNet validation dataset. Step One was executed for a total of 3000 queries unless the early stopping criterion was met. Step Two was run for an additional 2000 queries unless meeting the early stopping criterion. The main challenge for us was the CPU-based non-parallel PHA executables, which limited the number of images or queries we could process. In comparison, \cite{prokos2023squint} only utilized 30 images in experiments. 

For targeted hash-evasion attacks, experiment results are presented in Table \ref{tbl:simresult-31} and Appendix A.3.
With a focus on ASR(0.1), we observed that all targeted attacks failed, yielding ASR$(0.1)=0$. To our knowledge, no literature has ever reported non-zero ASR$(0.1)$, suggesting that all the PHAs are robust against blackbox targeted hash-evasion attacks. 



\begin{table}[t]
\centering
\caption{ASR($p$) of Targeted Hash-Evasion Attacks. ({\bf PH}: PhotoDNA. {\bf PD}: PDQ. {\bf NH}: NeuralHash)}
\label{tbl:simresult-31}
\begin{tabular}{|c|c|c|c|c|c|c|}
\hline
     &  &       &   &    & ZO-    & ZOsign  \\
     &  & SimBA &  & NES+  & sign- & -SGD\\
ASR  & SimBA  & +HSJA & NES & HSJA & SGD & +HSJA \\

\hline \hline

{\bf PH ASR(.1)} &     0\%&     0\%&     0\%&      0\%&   0\%&0\%\\


{\bf PD ASR(.1)} &     0\%&      0\%&      0\%&       0\%& 0\%&0\%\\


{\bf NH ASR(.1)} &     0\%&     0\%&     0\%&      0\%& 0\%&0\%\\

\hline \hline

\multicolumn{3}{|l|}{\cite{prokos2023squint} PH ASR(.3)  ($L_2=0.19$)}  & 83\% & \multicolumn{3}{|c|}{($2\times 10^6$ queries)} \\
\multicolumn{3}{|l|}{\cite{prokos2023squint} PD ASR(.3)  ($L_2=0.42$)} &  100\% & \multicolumn{3}{|c|}{($6\times 10^5$ queries)} \\
\hline

\end{tabular}
\end{table}













For untargeted attacks, experiment results are presented in Table \ref{tbl:simresult-21} and Appendix A.4.
With a focus on ASR(0.3), we see that PDQ was robust against all adversarial attacks, with a maximum ASR$(0.3)$ of 1\% only. It's worth noting that although \cite{jain2022adversarial} achieved a much higher attack success rate against PDQ, it required significantly more queries. The results also demonstrated that our defense method was effective because it drastically reduced ASR(0.3) and made the PHAs robust.

\begin{table}[t]
\centering
\caption{Adversary's ASR($p$) in Untargeted Blackbox Hash-Evasion Attacks, either without defense ($q=0$) or with defense $q=0.1$. ({\bf PH}: PhotoDNA. {\bf PD}: PDQ. {\bf NH}: NeuralHash. )}
\label{tbl:simresult-21}
\begin{tabular}{|c|c|c|c|c|c|c||   c|c|}
\hline
         &      &   & &  & ZO-       & ZOsign   & Defense  & Defense\\
      &   & SimBA &  & NES+  & sign- & -SGD     & +SimBA   & +NES  \\
ASR         &  SimBA   & +HSJA &  NES & HSJA & SGD & +HSJA    & +HSJA & +HSJA\\
\hline \hline

{\bf PH ASR(.3)} &     $1\%$&       \bf{92\%} &     $0\%$& $69\%$& $0\%$&$19\%$     &  27\%   & 21\% \\

{\bf PD ASR(.3)} &     0\%&       0\%&     0\%& {\bf 1\%} & 0\%&0\%    & 0\%   & 0\% \\

{\bf NH ASR(.3)} &     0\%&       14\%&     0\%& {\bf 28\%}   & 0\%&2\%   &  0\% & 2\% \\
\hline \hline

\multicolumn{9}{|l|}{\cite{prokos2023squint} PH ASR(.3)$=$ 90\%,   ($L_2=0.2$, HSJA with $3\times 10^4$ queries)} \\
\multicolumn{9}{|l|}{\cite{jain2022adversarial} PD ASR(.3)$=$ 73\%, ($L_2=0.1$, up to $8\times 10^6$ queries)} \\

\hline

\end{tabular}
\end{table}

\subsection{Experiment on Hash-Inversion Attack and Defense}

Experiments were conducted over three datasets: MNIST, CelebA, and STL-10. For each hashing algorithm and dataset, we trained the inversion network depicted in Fig. \ref{fig:inversion_model} for 50 epochs, with a batch size of 64 and an initial learning rate of 0.005. We utilized the MSE loss function with the AdamW optimizer and the cosine annealing learning rate scheduler. While MNIST and CelebA models were relatively easy to train, we encountered substantial challenges when training STL-10 models. We had to degrade STL-10 images to black-and-white in order to get meaningful reconstructed images. 

\begin{figure}[t]


\centerline{
\includegraphics[width=0.8\textwidth]{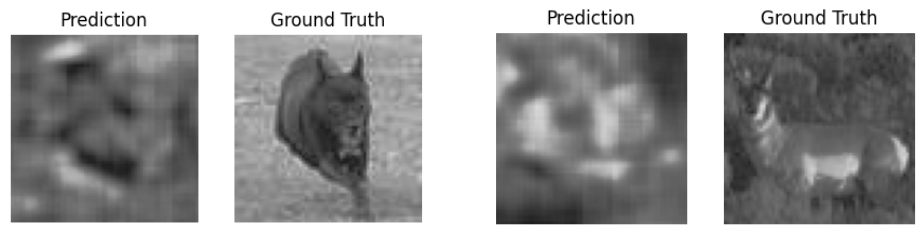}}
\centerline{\small (a) STL-10 image samples with NeuralHash, without defense}
\centerline{
\includegraphics[width=1.0\textwidth]{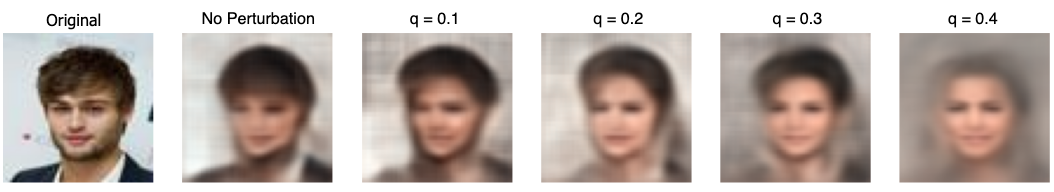}}
\centerline{
\includegraphics[width=1.0\textwidth]{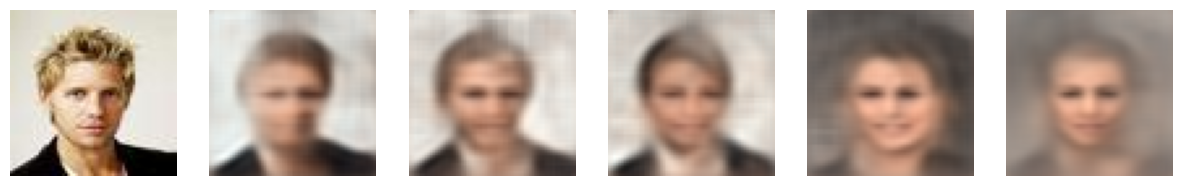}}
\centerline{\small (b) CelebA image samples with PhotoDNA, under various levels $q$ of defense}

\caption{Samples of true/original images and reconstructed images by the hash inversion algorithm.} 
\label{fig:simresult_52}
\end{figure}

Some sample images are shown in Fig. \ref{fig:simresult_52} whereas major experiment results are left to Appendix A.5 to save space. In general, hash-inversion failed over STL-10 while some level of success could be seen in MNIST and CelebA over PhotoDNA hash only. The reason is that the MNIST and CelebA images have similar or regular formation. The reconstructed images tend to converge to such common formation when the hash is long enough. However, even in this case, the reconstructed CelebA images could not be used to recognize the original face. 

The effectiveness of random hash perturbation defense can be clearly seen in Fig. \ref{fig:simresult_52}(b). Even the slightest perturbation $q=0.1$ made the reconstructed image quite different from the original image. 

\section{Conclusions}
\label{sec5}

This paper evaluates the security of three widely used perceptual hashing algorithms (PHAs): PhotoDNA, PDQ, and NeuralHash. Contrary to many existing studies, our findings show that these PHAs demonstrate notable robustness against both adversarial blackbox hash-evasion attacks and hash-inversion attacks. We attribute this robustness, in part, to the random variations inherent in the hash generation process. Building on this insight, we propose enhancing security by introducing additional randomization to the hash bits.

\bibliographystyle{IEEEbib}
\bibliography{reference.bib}

\newpage 
\appendix

\section{Appendix / supplemental material}

\subsection{Proof of Proposition 1 of Section 3.3}

Without loss of generality, consider the optimization problem (\ref{eq30}). A common approach involves utilizing $K$ randomly perturbed images to query the PHAs and then using the queried hashes to estimate the gradient as follows
\begin{equation}
  {\bf g} = \frac{1}{K} \sum_{k=1}^K \frac{1}{\beta} {\bf u}_k D_h({\bf h}_{\rm tar}, {\cal H}({\bf I}_0+\delta+\beta {\bf u}_k))
\end{equation}
where $\beta {\bf u}_k$ represents random perturbation \cite{ilyas2018black}. Firstly, due to the nature of PHAs, overly small perturbations may not induce any change in hash. Hence, relatively large $\beta$ is required, but this renders the gradient estimation unreliable. Secondly, the hash ${\cal H}({\bf I}_0 + \delta + \beta {\bf u}_k)$ randomly varies with the small perturbation $\beta {\bf u}_k$. The random variation can be modeled as a noise term $Z$ with variance $\sigma^2$. Following the analysis of \cite{aithal2022mitigating}, this noise is amplified by $1/\beta$, leading to the gradient estimate ${\bf g}$ being affected by a noise power $\sigma^2/\beta^2$, hence making it unreliable.

Due to the random nature of PHAs, it is almost impossible to limit the change of hash to just one or a few hash bits by perturbations. Practical $\sigma$ is usually around $0.1$. $\beta$ may be around $0.001$. As per \cite{aithal2022mitigating}, the signal-to-noise-ratio (SNR) of the gradient estimation can fall well below $-40$ dB, which makes the adversary's optimization not convergent or converge extremely slowly. This implies that PHAs are theoretically robust to blackbox attacks under the query budget constraint.

\subsection{Robustness Problem of PHAs to Normal Image Editing}

One of the special issue of PHAs is that they seem not robust to some normal imaging editing methods, for which \cite{steinebach2023analysis} examined PhotoDNA, while \cite{mckeown2023hamming} investigated PDQ and NeuralHash. Both studies demonstrated that PHAs were robust to certain edits like compression but vulnerable to others like cropping. While such prior works have already extensively investigated image editing on PHAs, a new evaluation is necessary due to ongoing algorithm updates. Notably, NeuralHash has received at least one update since its first release. This study addresses the gap in knowledge by evaluating the robustness of the updated algorithms and, more importantly, by providing the first comparison of the three PHAs.

Assume the original input to the PHA is an image ${\bf I}_0$, and the true output is an $N$-bit hash ${\bf h}_0={\cal H}({\bf I}_0)$, where ${\cal H}(\cdot)$ denotes the hashing algorithm.  The client adversary applies standard image editing operations over ${\bf I}_0$ to obtain an adversarial image ${\bf I}_{\rm adv}$ with a new $N$-bit hash ${\bf h}_{\rm adv}={\cal H}({\bf I}_{\rm adv})$.  A successful attack means ${\bf h}_{\rm adv}$ is sufficiently different from ${\bf h}_0$.


We experimented with four image editing operations that can keep the original high image quality: JPEG compression with random quality factors, resizing images with random scaling factors, filtering images with the vignette filter, or rotating images to random degrees. We calculated the hashes ${\bf h}_0$ and ${\bf h}_1$ of the images before and after the editing, and derived their distance $D_h({\bf h}_0, {\bf h}_1)$. Using 1000 images from the ImageNet validation dataset, we obtained ASR$(p)$ for various $p$ based on (\ref{eq80}), wherein the $L_2$ condition was removed by setting $\theta$ to infinity. 

Experiment results are presented in Tables \ref{tbl:simresult-11}. The data for ASR(.3) is highlighted in black because $p=0.3$ is the threshold we suggested to focus on. It is evident that all three PHAs were robust to compression and resizing, but not to filtering and rotation. 
In conjunction with the findings in \cite{mckeown2023hamming}\cite{struppek2022learning}, the PHAs are robust to many standard image editing operations except the following:
\begin{itemize}
    \item PhotoDNA is not robust to: filtering, rotating, resizing, mirroring, bordering, cropping; 
    \item PDQ is not robust to: filtering, rotating, mirroring \cite{mckeown2023hamming}, bordering \cite{mckeown2023hamming}, cropping \cite{mckeown2023hamming};
    \item NeuralHash is not robust to: filtering, rotating, mirroring \cite{mckeown2023hamming}.
\end{itemize}
NeuralHash was the most robust among the three. 
Surprisingly, the newest edition of NeuralHash that we experimented with was still not robust to filtering, rotating, and mirroring, similar to the first edition experimented in \cite{mckeown2023hamming}\cite{struppek2022learning}. Nevertheless, we believe this robustness problem can be mitigated by training the model to be invariant to such standard operations with data augmentations.



\begin{table}[h]
\centering
\caption{Adversary's ASR($p$) with Image Editing Attacks}
\label{tbl:simresult-11}
\begin{tabular}{|c|c|c|c|c|}
\hline
ASR      &  Compress & Resize  & Filter &  Rotate\\
\hline \hline

{PhotoDNA ASR(.1)} &     {3\%} &    99\%  &   $100\%$&     $100\%$\\
PhotoDNA ASR(.2) &     $0\%$&   55\%   &   $100\%$&  $100\%$\\
{\bf PhotoDNA ASR(.3)} &   {\bf 0\%} &    {\bf 0\%}  &    {\bf 94\%}&    {\bf 100\%}\\
PhotoDNA ASR(.4) &      $0\%$&    0\%  &   $37\%$&    $98\%$\\
\hline

{PDQ ASR(.1)} &     0\% &    27\%  &   {88\%}  &     100\%\\
PDQ ASR(.2) &     0\% &    0\%  &  67\%&   100\%\\
{\bf PDQ ASR(.3)} &     {\bf 0\%} &    {\bf 0\%}  &  {\bf 44\%} &   {\bf 100\%} \\
PDQ ASR(.4) &      0\% &   0\%  &   27\%&    100\%\\
\hline

{NeuralHash ASR(.1)} &  0\% &  {22\%}   &  96\% &     {100\%}\\
NeuralHash ASR(.2) &  0\% &  0\%    &   82\% &  100\%\\
{\bf NeuralHash ASR(.3)} &  {\bf 0\%} &  {\bf 0\%}    & {\bf 57\%} &   {\bf 96\%}\\
NeuralHash ASR(.4) &  0\% &  0\%    &  12\% &    31\%\\
\hline

\end{tabular}
\end{table}

\subsection{Extra Experiment Data of Targeted Hash-Evasion Attacks in Section 4.2}
Besides the data shown in Section 4.2 and Table \ref{tbl:simresult-31}, more experiment data of targeted attacks are presented in Table \ref{tbl:simresult-31-a}. We find that the robustness is extremely strong because of ASR$(p)\approx 0$ across all $p$. The only exception was NeuralHash, which exhibited ASR(0.4) of 43\% with SimBA. However, this isn't indicative of successful adversarial attacks; rather, it's due to numerous hashes generated by NeuralHash having $D_h({\bf h}_0, {\bf h}_1)$ as low as 0.3. Under NeuralHash, many images are inherently adversarial when evaluated at $p=0.4$.

Therefore, contrary to the prior claims that PHAs were not robust against targeted attacks \cite{prokos2023squint}, our findings demonstrate the strong robustness of all PHAs against targeted blackbox attacks. This is significant to the practical application of PHAs because the client adversary needs targeted attacks to ensure that its adversarial hashes do not inadvertently match illicit hashes in the database. The strong robustness means it is hard for the adversary to undertake either untargeted or targeted attacks without being caught.

\begin{table}[h]
\centering
\caption{Adversary's ASR($p$) in Targeted Blackbox Hash-Evasion Attacks. ({\bf PH}: PhotoDNA. {\bf PD}: PDQ. {\bf NH}: NeuralHash)}
\label{tbl:simresult-31-a}
\begin{tabular}{|c|c|c|c|c|c|c|}
\hline
     &  &       &   &    & ZO-    & ZOsign  \\
     &  & SimBA &  & NES+  & sign- & -SGD\\
ASR  & SimBA  & +HSJA & NES & HSJA & SGD & +HSJA \\

\hline \hline

{\bf PH ASR(.1)} &     0\%&     0\%&     0\%&      0\%&   0\%&0\%\\
PH ASR(.2) &     0\%&      0\%&      0\%&   0\%& 0\%&0\%\\
PH ASR(.3) &     0\%&       0\%&     0\%&    0\%& 0\%&0\%\\
PH ASR(.4) &      1\%&      {\bf 2\%}&      0\%&     0\%& 0\%&0\%\\
\hline

{\bf PD ASR(.1)} &     0\%&      0\%&      0\%&       0\%& 0\%&0\%\\
PD ASR(.2) &     0\%&      0\%&      0\%&   0\%& 0\%&0\%\\
PD ASR(.3) &     0\%&       0\%&     0\%&    0\%& 0\%&0\%\\
PD ASR(.4) &      0\%&      0\%&      0\%&     0\%& 0\%&0\%\\
\hline

{\bf NH ASR(.1)} &     0\%&     0\%&     0\%&      0\%& 0\%&0\%\\
NH ASR(.2) &     0\%&      0\%&      0\%&   0\%& 0\%&0\%\\
NH ASR(.3) &     0\%&       0\%&     0\%&    0\%& 0\%&0\%\\
NH ASR(.4) &      {\bf 43\%}&      11\%&      0\%&     0\%& 0\%&0\%\\
\hline \hline

\multicolumn{3}{|l|}{\cite{prokos2023squint} PH ASR(.3)  ($L_2=0.19$)}  & 83\% & \multicolumn{3}{|c|}{($2\times 10^6$ queries)} \\
\multicolumn{3}{|l|}{\cite{prokos2023squint} PD ASR(.3)  ($L_2=0.42$)} &  100\% & \multicolumn{3}{|c|}{($6\times 10^5$ queries)} \\
\hline

\end{tabular}
\end{table}

\subsection{Extra Experiment Data of Untargeted Blackbox Hash-Evasion Attacks in Section 4.2}

\subsubsection{Experiment data and sample images of untargeted attacks without defense}
\begin{figure}[h]
\centerline{
\includegraphics[width=0.3\textwidth]{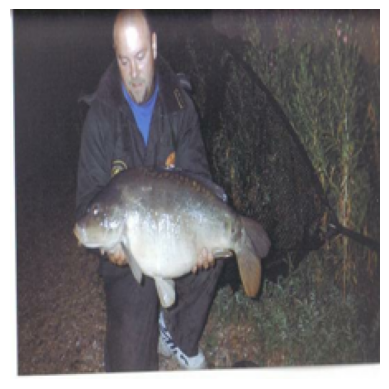}}
\centerline{\small (a) Original image ${\bf I}_0$.}
\centerline{
\includegraphics[width=0.8\textwidth]{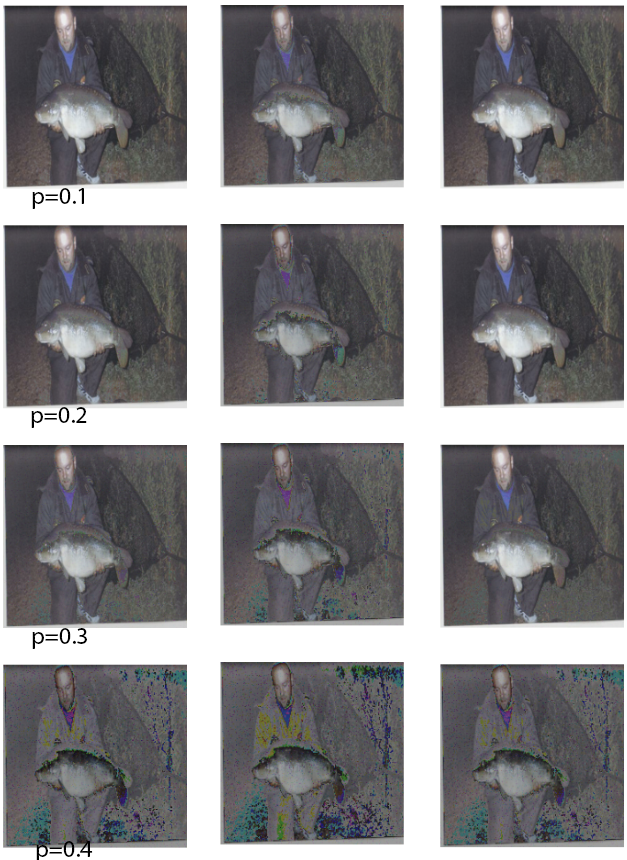}}
\centerline{\small PhotoDNA  \hspace{2.5cm} PDQ  \hspace{2.5cm} NeuralHash}
\centerline{\small (b) Adversarial images ${\bf I}_{\rm adv}$.}

\caption{Samples of an original image and the adversarial images created by the proposed untargeted blackbox attack algorithm JSHA (NES+HSJA).} 
\label{fig:simfig_21-a}
\end{figure}

\begin{figure}[h]
\centerline{
\includegraphics[width=0.8\textwidth]{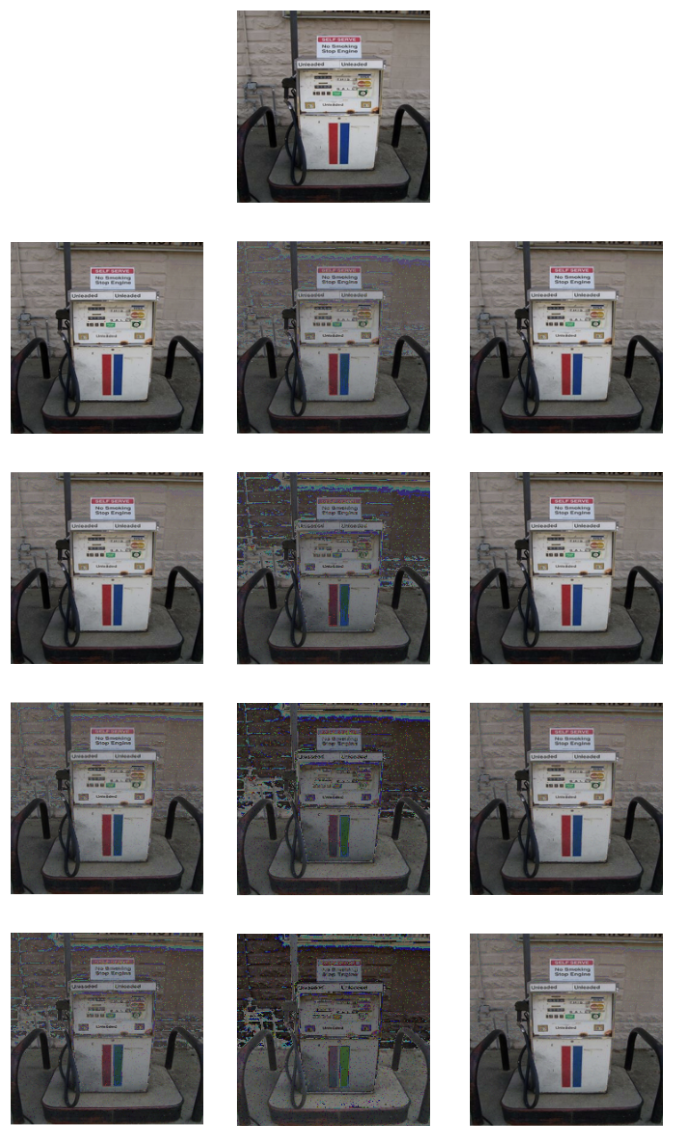}}
\centerline{\small PhotoDNA  \hspace{2.5cm} PDQ  \hspace{2.5cm} NeuralHash}

\caption{Samples of an original image and the adversarial images created by the proposed untargeted blackbox attack algorithm JSHA (NES+HSJA).} 
\label{fig:simfig_21-b}
\end{figure}

Besides the experiment data shown in Section 4.2 and Table \ref{tbl:simresult-21}, extra experiment data are presented in Table \ref{tbl:simresult-21-a}. 
Sample adversarial images generated with our proposed JSHA with various pre-set threshold $p$ are depicted in Fig. \ref{fig:simfig_21-a} and Fig. \ref{fig:simfig_21-b}, with their corresponding distortion metrics listed in Table \ref{tbl:simresult-23-a}. Notably, the images with $p=0.4$ were deemed invalid adversarial images as their distortion was not under the threshold $\theta=0.1$, resulting in extremely noisy images. Similarly, 
for PDQ, the adversarial images with $p=0.2$ and $0.3$ were also invalid. 

Therefore, contrary to previous studies suggesting that PHAs were not robust against untargeted attacks, we found that some PHAs like PDQ and NeuralHash could be considered as robust if realistic constraints regarding image distortion and query budget are taken into account.  

\begin{table}[h]
\centering
\caption{Adversary's ASR($p$) in Untargeted Blackbox Hash-Evasion Attacks. ({\bf PH}: PhotoDNA. {\bf PD}: PDQ. {\bf NH}: NeuralHash. )}
\label{tbl:simresult-21-a}
\begin{tabular}{|c|c|c|c|c|c|c|}
\hline
         &      &   & &  & ZO-       & ZOsign  \\
      &   & SimBA &  & NES+  & sign- & -SGD\\
ASR         &  SimBA   & +HSJA &  NES & HSJA & SGD & +HSJA \\
\hline \hline

PH ASR(.1) &     $100\%$&     $100\%$&     $18\%$& $100\%$& $1\%$&$98\%$\\
PH ASR(.2) &     $80\%$&      $100\%$&      $2\%$& $98\%$& $0\%$&$90\%$\\
{\bf PH ASR(.3)} &     $1\%$&       \bf{92\%} &     $0\%$& $69\%$& $0\%$&$19\%$\\
PH ASR(.4) &      $0\%$&      $0\%$&      $0\%$& $1\%$
& $0\%$&$0\%$\\
\hline 

PD ASR(.1) &     0\%&     45\%&     0\%& 64\%& 0\%&8\%\\
PD ASR(.2) &     0\%&      1\%&      0\%& 10\%& 0\%&0\%\\
{\bf PD ASR(.3)} &     0\%&       0\%&     0\%& {\bf 1\%} & 0\%&0\%\\
PD ASR(.4) &      0\%&      0\%&      0\%& 0\%& 0\%&0\%\\
\hline

NH ASR(.1) &     4\%&  88\%&   0\%&99\%& 0\%&94\%\\
NH ASR(.2) &     2\%&      23\%&      0\%&77\%   & 0\%&42\%\\
{\bf NH ASR(.3)} &     0\%&       14\%&     0\%& {\bf 28\%}   & 0\%&2\%\\
NH ASR(.4) &      0\%&      0\%&     0\%&0\%  & 0\%&0\%\\
\hline \hline

\multicolumn{7}{|l|}{\cite{prokos2023squint} PH ASR(.3)$=$ 90\%,   ($L_2=0.2$, HSJA with $3\times 10^4$ queries)} \\
\multicolumn{7}{|l|}{\cite{jain2022adversarial} PD ASR(.3)$=$ 73\%, ($L_2=0.1$, up to $8\times 10^6$ queries)} \\

\hline

\end{tabular}
\end{table}

\begin{table}[h]
\centering
\caption{Distortion of the Adversarial Image Samples in Fig. \ref{fig:simfig_21-a}(b).}
\label{tbl:simresult-23-a}

\begin{tabular}{|c|c|| c|c|c|}
\hline 
Measure & Threshold $p$ &  PhotoDNA & PDQ  &  NeuralHash \\
\hline 
    & $p=0.1$ & 0.02 & 0.07 & 0.02 \\
$L_2$ & $p=0.2$ & 0.02 & {\bf 0.11} & 0.04 \\
     & $p=0.3$ & 0.05 & {\bf 0.14} & 0.07  \\
     & $p=0.4$ & {\bf 0.17} & {\bf 0.19} & {\bf 0.19} \\
\hline

\end{tabular}
\end{table}

\subsubsection{Experiment data of untargeted attacks under proposed defense}

Observing that some untargeted attack algorithms such as SimBA+HSJA achieved relataively high ASR(0.3) over some PHAs such as PhotoDNA, we experimented with the hash perturbation defense method proposed in Section 3.3. 
The level of our random hash perturbation was denoted as $q$, where   $0\le q \le1$, which stand for the ratio of the hash bits that were randomly flipped. 

Besides the experiment results presented in Section 4.2 and Table \ref{tbl:simresult-21}, more experiment data are presented in Tables \ref{tbl:simresult-21-e},  \ref{tbl:simresult-21-f}, and \ref{tbl:simresult-21-g}. It can be seen that random perturbation reduced ASR($p$) drastically, which means the attacks became much less successful or even failed. Even a slight perturbation with $q=0.1$ could prevent most of the attacks.
\begin{table}[h]
\centering
\caption{Adversary's ASR($0.1$) in Untargeted Blackbox Hash-Evasion Attacks when Defense was Applied. ({\bf PH}: PhotoDNA. {\bf PD}: PDQ. {\bf NH}: NeuralHash. )}
\label{tbl:simresult-21-e}
\begin{tabular}{|c|c|c|c|}
\hline
         &   &  & ZOsign  \\
      & SimBA & NES+  & -SGD\\
ASR         & +HSJA & HSJA & +HSJA \\
\hline \hline

PH q =0.1&     100\%& $100\%$&$100\%$\\
PH  q =0.2&      100\%& $100\%$&$100\%$\\
PH q =0.3&       100\%& $100\%$&$100\%$\\
\hline 

PD q =0.1&     0\%& 0\%&0\%\\
PD q =0.2&      0\%& 0\%&0\%\\
PD q =0.3&       0\%& 0\%&0\%\\
\hline

NH  q =0.1&  21\%&18\%&15\%\\
NH q =0.2&      20\%&18\%&15\%\\
NH q =0.3&       21\%& 18\%&15\%\\
\hline

\end{tabular}
\end{table}
\begin{table}[h]
\centering
\caption{Adversary's ASR($0.2$) in Untargeted Blackbox Hash-Evasion Attacks when Defense was Applied. ({\bf PH}: PhotoDNA. {\bf PD}: PDQ. {\bf NH}: NeuralHash. )}
\label{tbl:simresult-21-f}
\begin{tabular}{|c|c|c|c|}
\hline
         &   &  & ZOsign  \\
      & SimBA & NES+  & -SGD\\
ASR         & +HSJA & HSJA & +HSJA \\
\hline \hline

PH q =0.1&     98\%& 99\%&95\%\\
PH  q =0.2&      98\%& 99\%&95\%\\
PH q =0.3&       98\%& 99\%&95\%\\
\hline 

PD q =0.1&     0\%& 0\%&0\%\\
PD q =0.2&      0\%& 0\%&0\%\\
PD q =0.3&       0\%& 0\%&0\%\\
\hline

NH  q =0.1&  5\%&5\%&2\%\\
NH q =0.2&      5\%&6\%&3\%\\
NH q =0.3&       5\%& 6\%&2\%\\
\hline

\end{tabular}
\end{table}
\begin{table}[h]
\centering
\caption{Adversary's ASR($0.3$) in Untargeted Blackbox Hash-Evasion Attacks when Defense was Applied. ({\bf PH}: PhotoDNA. {\bf PD}: PDQ. {\bf NH}: NeuralHash. )}
\label{tbl:simresult-21-g}
\begin{tabular}{|c|c|c|c|}
\hline
         &   &  & ZOsign  \\
      & SimBA & NES+  & -SGD\\
ASR         & +HSJA & HSJA & +HSJA \\
\hline \hline

PH q =0.1&     27\%& 21\%&5\%\\
PH  q =0.2&      27\%& 21\%&5\%\\
PH q =0.3&       27\%& 21\%&5\%\\
\hline 

PD q =0.1&     0& 0\%&0\%\\
PD q =0.2&      0\%& 0\%&0\%\\
PD q =0.3&       0\%& 0\%&0\%\\
\hline

NH  q =0.1&  0\%&2\%&0\%\\
NH q =0.2&      0\%&2\%&0\%\\
NH q =0.3&       0\%& 0\%&0\%\\
\hline

\end{tabular}
\end{table}

\subsection{Extra Experiment Results of Hash-Inversion Attacks in Section 4.3}

\subsubsection{Experiment data and sample images of hash-inversion attacks without defense}
\begin{figure}[h]
\centerline{
\includegraphics[width=0.6\textwidth]{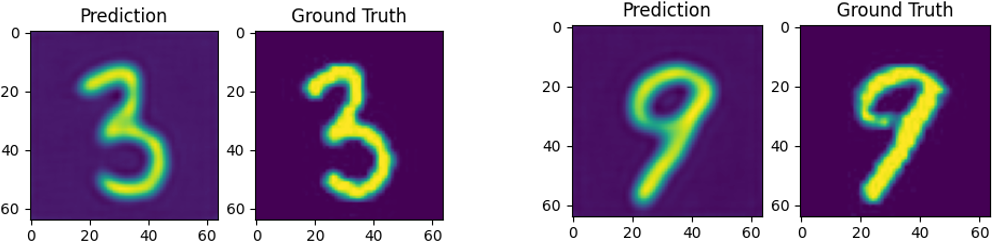}}
\centerline{\small (a) PhotoDNA. Left: (.14, .83, .05). Right: (.12, .86, .07)}
\centerline{
\includegraphics[width=0.6\textwidth]{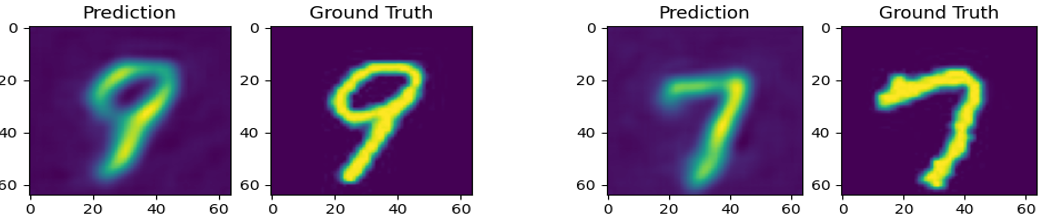}}
\centerline{\small (b) NeuralHash. Left: (.26, .61, .10). Right: (.34, .52, .14)}
\caption{Samples of true MNIST images and adversarial images generated by hash inversion attacks based on hashes of (a) PhotoDNA, and (b) NeuralHash. The numbers are ($L_2$, SSIM, LPIPS) measures.} 
\label{fig:simresult_51-a}
\end{figure}

\begin{figure}[h]
\centerline{
\includegraphics[width=0.8\textwidth]{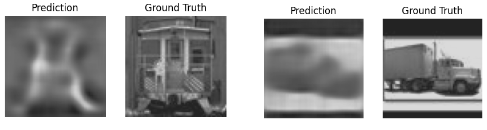}}
\centerline{
\includegraphics[width=0.8\textwidth]{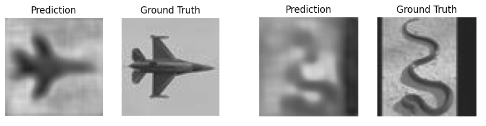}}
\centerline{\small (a) Using PhotoDNA Hash bits}
\centerline{
\includegraphics[width=0.8\textwidth]{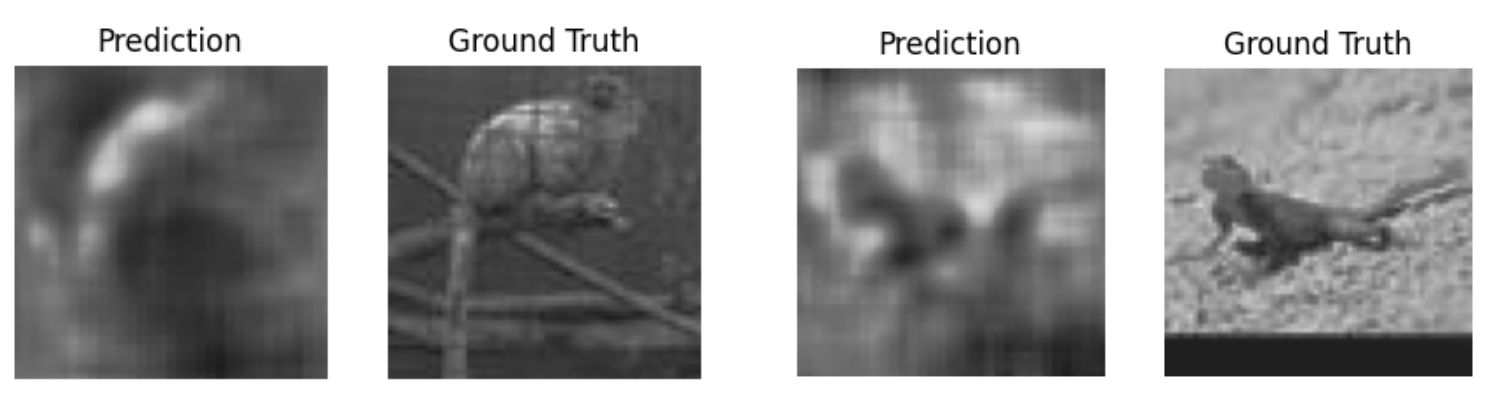}}
\centerline{\small (b) Using PDQ Hash bits}
\centerline{
\includegraphics[width=0.8\textwidth]{fig_used/neuralhash1.png}}
\centerline{\small (c) Using NeuralHash hash bits}
\caption{Samples of true STL-10 images and adversarial images generated by hash inversion attacks.} 
\label{fig:simresult_69}
\end{figure}

\begin{figure}[h]
\centerline{
\includegraphics[width=0.6\textwidth]{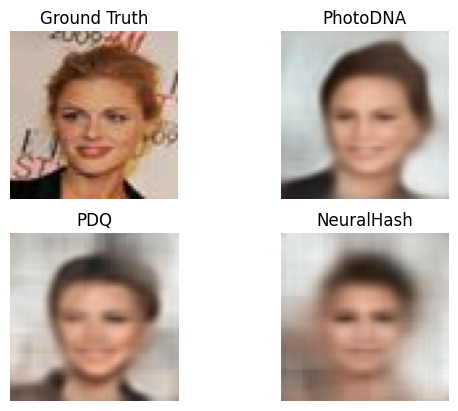}}
\centerline{\small (a) Our model. \{$L_2$, SSIM, LPIPS\} $=$ \{{\bf .14}, {\bf .65}, {\bf .34}\} (PhotoDNA), }
\centerline{\small $=$ \{.15, .60, .36\} (PDQ), \{.20, .40, .50\} (NeuralHash)}

\centerline{
\includegraphics[width=0.6\textwidth]{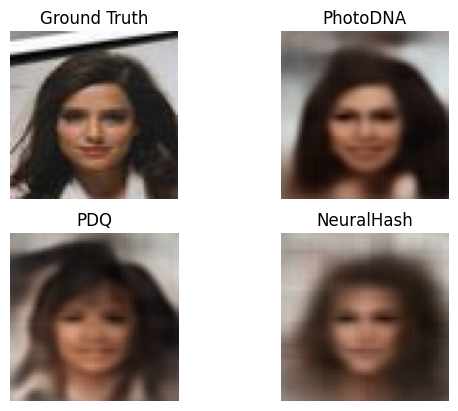}}
\centerline{\small (b) Our model. \{$L_2$, SSIM, LPIPS\} $=$ \{{\bf .17}, {\bf .63}, {\bf .34}\} (PhotoDNA), }
\centerline{\small $=$ \{.21, .50, .34\} (PDQ), \{.20, .47, .45\} (NeuralHash)}

\centerline{
\includegraphics[width=0.8\textwidth]{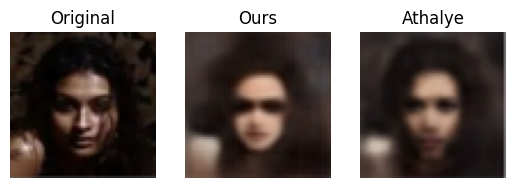}}
\centerline{\small (c) Compare images generated by our model and \cite{Athalye2021inverting}.}


\caption{Samples of true CelebA images and reconstructed images generated by hash inversion attacks from hashes of PhotoDNA, PDQ, and NeuralHash. The numbers are quality metrics \{$L_2$, SSIM, LPIPS\} of reconstructed images.} 
\label{fig:simresult_51-c}
\end{figure}

Besides the experiment sample images presented in Section 4.3 and Fig. \ref{fig:simresult_52}, a complete set of experiment data of image reconstruction quality are in Table \ref{tbl:simresult-4-a}. In addition, Fig. \ref{fig:simresult_51-a} shows sample images of MNIST dataset reconstruction, Fig. \ref{fig:simresult_69} shows sample images of STL-10 dataset reconstruction, whereas Fig. \ref{fig:simresult_51-c} shows sample images of CelebA dataset reconstruction. All these experiment results were obtained without the defense being applied.

Generally, the hash-inversion attacks worked with some success when there was a lot of regularity in the dataset (e.g., MNIST and CelebA). In this case, the hash-inversion algorithm gave an image of such regularity only. The longer hash values leaked more information and allowed better hash-inversion. Nevertheless, even in the best case, it was still impossible to match the reconstructed human faces with the true faces in the CelebA images. 

When the dataset was diverse (e.g., STL-10), we were not able to learn any meaningful inversions. This demonstrated that the PHAs were robust against hash-inversion attacks in practical applications. In addition, PHAs with smaller number of output bits were always preferable because natural images were not very standardized, so the smaller number of bits could help to prevent information leakage/hash inversion. 

The reason for \cite{Athalye2021inverting} to draw the false conclusion of successful hash-inversion was because only CelebA images were used over the long PhotoDNA hashes, and the authors overlooked the fact that the faces can not be discriminated based on the reconstructed images.


\begin{table}[h]
\centering
\caption{Quality of Adversarial Images Created in Hash Inversion Attacks ({\bf PH}: PhotoDNA. {\bf PD}: PDQ. {\bf NH}: NeuralHash)}
\label{tbl:simresult-4-a}
\begin{tabular}{|c|c|c|c|c|}
\hline
 Dataset & Hash &  $L_2$  & SSIM  & LPIPS  \\
 \hline \hline
  & PH & $\mathbf{0.08}\pm0.02$&      $\mathbf{0.79}\pm0.02$& $\mathbf{0.18}\pm0.03$\\
  MNIST & PD &  - & - & - \\
  & NH & $0.19\pm0.05$&     $0.35\pm0.10$& $0.33\pm0.08$\\
\hline \hline
 & PH & $\mathbf{0.13}\pm0.03$&     $\mathbf{0.57}\pm0.08$&     $\mathbf{0.40}\pm0.07$\\
 CelebA & PD & $0.23\pm0.07$&     $0.44\pm0.07$&     $0.43\pm0.05$\\
 & NH & $0.23\pm0.05$&     $0.29\pm0.09$&     $0.53\pm0.07$\\
\hline \hline
 & PH & $\mathbf{0.14}\pm0.03$&     $\mathbf{0.41}\pm0.09$&     $\mathbf{0.56}\pm0.06$\\
 STL-10 & PD & $0.24\pm0.05$&     $0.26\pm0.05$&     $0.65\pm0.02$\\
 & NH & $0.23\pm0.05$&     $0.14\pm0.09$&     $0.67\pm0.07$\\
\hline   
\end{tabular}
\end{table}







\subsubsection{Experiment data hash-inversion attacks under the proposed defense}

To mitigate the potential hash inversion attacks over regular images, we experimented with the proposed random hash perturbation defense algorithm. Some sample images were shown in Fig. \ref{fig:simresult_52}(b) and explained briefly in Section 4.3. Extra experiment data are shown in Tables \ref{tbl:simresult-40-a}, \ref{tbl:simresult-40-b} and \ref{tbl:simresult-40-c}.

Compare these data with those in Table \ref{tbl:simresult-4-a}, it is clear that as we increased the level of perturbation to the hash bits, the quality of reconstructed images decreased as expected. Therefore, the random perturbation defense does help to defend against hash-inversion attacks.

\begin{table}[h]
\centering
\caption{Quality of Inverted Images when Defended with q = 0.1 ({\bf PH}: PhotoDNA. {\bf PD}: PDQ. {\bf NH}: NeuralHash)}
\label{tbl:simresult-40-a}
\begin{tabular}{|c|c|c|c|c|}
\hline
 Dataset & Hash &  $L_2$  & SSIM  & LPIPS  \\
\hline \hline
 & PH & $0.14\pm0.03$&     $0.41\pm0.00$&     $0.56\pm0.06$\\
 STL-10 & PD & $0.24\pm0.05$&     $0.26\pm0.05$&     $0.65\pm0.02$\\
 & NH & $0.23\pm0.05$&     $0.14\pm0.07$&     $0.67\pm0.03$\\
\hline   
\end{tabular}
\end{table}

\begin{table}[h]
\centering
\caption{Quality of Inverted Images when Defended with q = 0.2 ({\bf PH}: PhotoDNA. {\bf PD}: PDQ. {\bf NH}: NeuralHash)}
\label{tbl:simresult-40-b}
\begin{tabular}{|c|c|c|c|c|}
\hline
 Dataset & Hash &  $L_2$  & SSIM  & LPIPS  \\
\hline \hline
 & PH & $0.18\pm0.04$&     $0.26\pm0.08$&     $0.68\pm0.05$\\
 STL-10 & PD & $0.23\pm0.05$&     $0.24\pm0.08$&     $0.65\pm0.04$\\
 & NH & $0.23\pm0.05$&     $0.17\pm0.08$&     $0.77\pm0.07$\\
\hline   
\end{tabular}
\end{table}

\begin{table}[h]
\centering
\caption{Quality of Inverted Images when Defended with q = 0.3 ({\bf PH}: PhotoDNA. {\bf PD}: PDQ. {\bf NH}: NeuralHash)}
\label{tbl:simresult-40-c}
\begin{tabular}{|c|c|c|c|c|}
\hline
 Dataset & Hash &  $L_2$  & SSIM  & LPIPS  \\
\hline \hline
 & PH & $0.20\pm0.06$&     $0.21\pm0.09$&     $0.71\pm0.05$\\
 STL-10 & PD & $0.23\pm0.06$&     $0.20\pm0.08$&     $0.68\pm0.04$\\
 & NH & $0.23\pm0.05$&     $0.17\pm0.08$&     $0.78\pm0.04$\\
\hline   
\end{tabular}
\end{table}

\end{document}